 \documentclass[twocolumn,PRL,aps,psfig,preprintnumbers,superscriptaddress]{revtex4}
\usepackage{graphicx}
\usepackage{epstopdf}
\usepackage{float}
\usepackage{color}
\usepackage{amsmath}

\begin{document}

\title{Possible Coexistence of Antiferromagnetic and Ferromagnetic Spin Fluctuations in the Spin-triplet Superconductor UTe$_2$ Revealed by $^{125}$Te  NMR under Pressure }
\author{Devi V. Ambika}
\affiliation{Ames Laboratory, U.S. DOE, Ames, Iowa 50011, USA}
\affiliation{Department of Physics and Astronomy, Iowa State University, Ames, Iowa 50011, USA}
\author{Qing-Ping Ding}
\affiliation{Ames Laboratory, U.S. DOE, Ames, Iowa 50011, USA}
 \affiliation{Department of Physics and Astronomy, Iowa State University, Ames, Iowa 50011, USA}
\author{Khusboo Rana}
\affiliation{Ames Laboratory, U.S. DOE, Ames, Iowa 50011, USA}
\affiliation{Department of Physics and Astronomy, Iowa State University, Ames, Iowa 50011, USA}
\author{Corey E. Frank}
\affiliation{NIST Center for Neutron Research, National Institute of Standards and Technology, Gaithersburg, MD 20899, USA}
\affiliation{Center for Nanophysics and Advanced Materials, Department of Physics, University of Maryland, College Park, MD 20742, USA}
\author{Elizabeth L. Green}
\affiliation{National High Magnetic Field Laboratory, Florida State University, Tallahassee, Florida 32310, USA}
\author{Sheng Ran$\footnote[1]{Present address: Department of Physics, Washington University,  St. Louis, MO}$}
\affiliation{NIST Center for Neutron Research, National Institute of Standards and Technology, Gaithersburg, MD 20899, USA}
\affiliation{Center for Nanophysics and Advanced Materials, Department of Physics, University of Maryland, College Park, MD 20742, USA}
\author{Nicholas P. Butch}
\affiliation{NIST Center for Neutron Research, National Institute of Standards and Technology, Gaithersburg, MD 20899, USA}
\affiliation{Center for Nanophysics and Advanced Materials, Department of Physics, University of Maryland, College Park, MD 20742, USA}
\author{Yuji Furukawa}
\affiliation{Ames Laboratory, U.S. DOE, Ames, Iowa 50011, USA}
\affiliation{Department of Physics and Astronomy, Iowa State University, Ames, Iowa 50011, USA}

\date{\today}

\begin{abstract} 

    A spin-triplet superconducting state mediated by ferromagnetic (FM) spin fluctuations has been suggested to occur in the newly discovered heavy-fermion superconductor UTe$_2$.     However, the recent neutron scattering measurements revealed the presence of antiferromagnetic (AFM) spin fluctuations in UTe$_2$.   Here, we report the $^{125}$Te nuclear magnetic resonance (NMR) studies of a single-crystal UTe$_2$, suggesting the coexistence of FM and AFM spin fluctuations  in UTe$_2$.     Owing to the two different Te sites in the compound, we conclude that the FM spin fluctuations are dominant within ladders and the AFM spin fluctuations originate from the inter-ladder magnetic coupling.     Although AFM spin fluctuations exist in the system, the FM spin fluctuations in the ladders may play an important role in the appearance of the spin-triplet superconducting state of UTe$_2$.

\end{abstract}

\maketitle


    The interplay between magnetic fluctuations  and superconductivity (SC) is one of the central issues in unconventional superconductors  such as iron-based superconductors, high $T_{\rm c}$ cuprates, and heavy fermions.
    UTe$_2$ is a newly discovered heavy-fermion superconductor with a superconducting transition temperature $T_{\rm c}$ $\sim$ 1.6--2.0 K \cite{Ran2019,Aoki2019,Rosa2021} and was proposed to be located at the end member of U-based ferromagnetic (FM) superconductors.
      Distinct from the previously discovered FM superconductors such as UGe$_2$, UCoGe, and URhGe which exhibit long-range FM order \cite{Aoki20192}, UTe$_2$ does not show any magnetic order down to 0.25 K \cite{Ran2019, Aoki2019, Ran2019_Nat, Sundar2019}, making it unique in the family of U-based FM superconductors.
    The observation of the nature of unconventional spin-triplet SC in UTe$_2$, such as very anisotropic SC upper critical field $H_{\rm c2}$ exceeding the Pauli limit \cite{Ran2019,Aoki2019,Knebel2019}, multiple SC phases \cite{Braithwaite2019,Hayes2021,Thomas2020}, and time-reversal symmetry breaking \cite{Hayes2021,Wei2022}, has also sparked a large volume of research activity on the compound \cite{Aoki20192}.

    Initially, FM spin fluctuations have been understandably considered to play an important role for the triplet paring, as suggested by muon spin relaxation \cite{Sundar2019} and nuclear magnetic resonance (NMR) measurements \cite{Tokunaga2019}.   
   On the other hand, recently, antiferromagnetic (AFM) spin fluctuations with the incommensurate wave-vector of $q$ = (0, 0.57, 0) have been detected by neutron scattering (NS) measurements \cite{Knafo2021, Duan2020}. 
   Those results indicate that  the nature of the magnetic fluctuations in UTe$_2$ are complicated and still under debate. 

     The nature of the magnetically ordered state as well as the SC phases  under pressure  is also still an open question. 
     With the application of pressure,  two SC phases appear around 0.25 GPa \cite{Thomas2020}. 
    While the SC phase with lower $T_{\rm c}$ (SC1) is suppressed continuously with pressure, the SC phase with higher $T_{\rm c}$ (SC2) is enhanced and takes a maximum $T_{\rm c}$ $\sim$ 3 K at around 1.2 GPa, and is suppressed rapidly at higher pressures. 
     Above the critical pressure $p_{\rm c}$~$\sim$~1.5~GPa, a magnetic phase appears \cite{Ran_PRB2020,Braithwaite2019,Thomas2020,Li2021}. 
      Initially,  a FM ordered state was suggested  for the pressure-induced magnetically ordered state \cite{Ran_PRB2020,Braithwaite2019}, however, recent studies proposed an AFM state \cite{Thomas2020,Li2021}. 

   NMR  is a powerful technique to investigate low-energy spin fluctuations and superconducting properties  from a microscopic point of view.
  The temperature ($T$) dependence of the nuclear spin-lattice relaxation rate (1/$T_1$) reflects the wave vector $q$-summed dynamical susceptibility at nuclear sites.
  On the other hand, NMR spectrum measurements,  in particular, the Knight shift $K$, give us information on local static magnetic properties. 
  Furthermore, in the SC state, the temperature dependences of $K$ and 1/$T_1$ provide important information about the SC gap structure. 
In fact, recent NMR measurements on single crystals revealed strong and slow spin fluctuations in the normal state \cite{Tokunaga2019, Tokunaga2022} and also provided key experimental results supporting the spin-triplet superconducting state in UTe$_2$ \cite {Nagamine2019,Nagamine2021,Fujibayashi2022}.

    In this paper, we have carried out $^{125}$Te NMR measurements using a $^{125}$Te-enriched single crystal of UTe$_2$ to investigate the evolution of magnetic fluctuations in UTe$_2$ under pressure. 
   The most striking result obtained here is the observation of the possible coexistence of AFM and FM spin  fluctuations in UTe$_2$.
   Owing to the two different Te sites in the compound, the FM spin fluctuations are considered  to be dominated within the ladders while the AFM spin fluctuations originate from the inter-ladder magnetic couplings. 
    The observed  results are consistent with the recent NS data where intra-ladder and inter-ladder magnetic couplings are ferromagnetic and antiferromagnetic, respectively \cite{Knafo2021, Duan2020}. 


\begin{figure}[tb]
\includegraphics[width=\columnwidth]{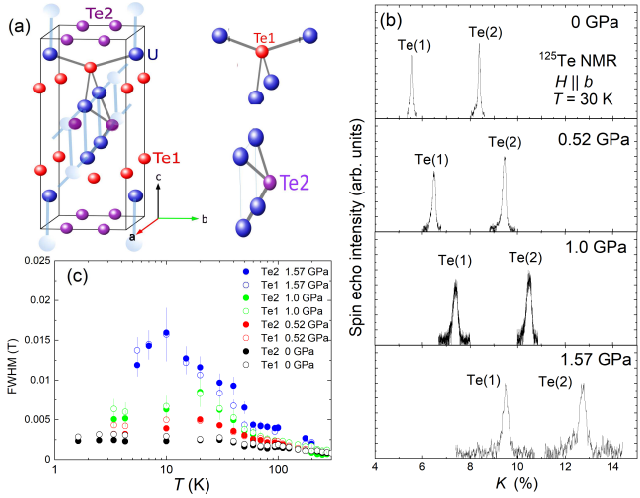} 
\caption{(a) Crystal structure of UTe$_2$.  
    The U atoms (blue circles) form a two-leg ladder structure. 
    The light blue circles representing the U atom outside the unit cell are shown to emphasize the ladder structure. 
    The local environment of Te1 and Te2 sites are shown on the right. 
(b) Pressure dependence of the $H$-swept $^{125}$Te-NMR spectra of a single crystal  UTe$_2$ at $T$ = 30 K for $H$ $\parallel$ $b$ where the horizontal axis is Knight shift $K$ defined by $K$ = ($H_0$ -- $H$)/$H$ where $H_0$ = 2$\pi$$f$/$\gamma_{\rm N}$,  $f$ is NMR resonance frequency, and $H$ is the external magnetic field. 
   For $p$ = 0, 0.52 and 1.0 GPa, the spectra were measured at $f$~=~41~MHz ($H_0$~=~3.0472~T). $f$~=~45.35~MHz ($H_0$~=~3.3717~T) was used at $p$ = 1.57 GPa.   
(c)  $T$ and $p$ dependences of FWHMs of Te1 and Te2 sites estimated from $^{125}$Te-NMR spectra.  
}
\label{fig:Fig1}
\end{figure}

        The $^{125}$Te-enriched single crystal  (3 $\times$ 1 $\times$ 0.2 mm$^3$)  of   UTe$_2$ with $T_{\rm c}$ = 1.6 K at zero magnetic field ($H$)  at ambient pressure was synthesized by chemical vapor transport method using iodine as the transport agent \cite{Ran2019}.
      NMR measurements of $^{125}$Te ($I$ = $\frac{1}{2}$, $\frac{\gamma_{\rm N}}{2\pi}$ = 13.454 MHz/T) nuclei were conducted using a laboratory-built phase-coherent spin-echo pulse spectrometer up to a pressure of 1.57 GPa with a NiCrAl/CuBe piston-cylinder cell using  Daphne 7373 as the pressure transmitting medium. 
    Pressure calibration was accomplished by $^{63}$Cu nuclear quadrupole resonance in Cu$_2$O \cite{Fukazawa2007,Reyes1992} at 77 K.
   $T_{\rm c}$ = 2.3~K for 0.52 GPa and 2.8 K for 1.0 GPa at $H$ = 0 were determined by in-situ AC susceptibility measurements using an NMR tank circuit. 
    The values of $T_{\rm c}$ are consistent with previous papers \cite{Ran_PRB2020,Thomas2020}.  
     The $^{125}$Te-NMR spectra were obtained by sweeping $H$ at fixed NMR frequencies ($f$) where $H$ was applied parallel to the $b$ axis.
     The 1/$T_{\rm 1}$ was measured with a saturation recovery method \cite{T1}.

      UTe$_2$ crystallizes in a body-centered orthorhombic structure with the $Immm$ space group \cite{Hutanu2020}  where the U atoms form a two-leg ladder structure with legs along $a$ axis and rung along the $c$ axis as shown in Fig.~\ref{fig:Fig1}(a) \cite{VESTA}. 
    There are two crystallographic inequivalent Te sites occupying $4j$ and $4h$ sites with point symmetries $mm2$ and $m2m$, respectively. 
     Following the previous paper \cite{Tokunaga2019}, these sites are denoted by Te1 and Te2. 
   Te1 is located inside the distorted tetrahedron formed by the first and second nearest-neighbors four U atoms which are belonging to three ladders.
On the other hand, Te2 is surrounded by the four nearest neighbor U atoms forming a square-like structure within a ladder.    
    Thus NMR measurements for Te2 mainly pick up the local magnetic properties of each ladder while Te1 NMR provides the local information of magnetic properties related to inter-ladder coupling. 
    Those differences in the environments for the Te sites are important as discussed below.


\begin{figure}[tb]
\includegraphics[width=\columnwidth]{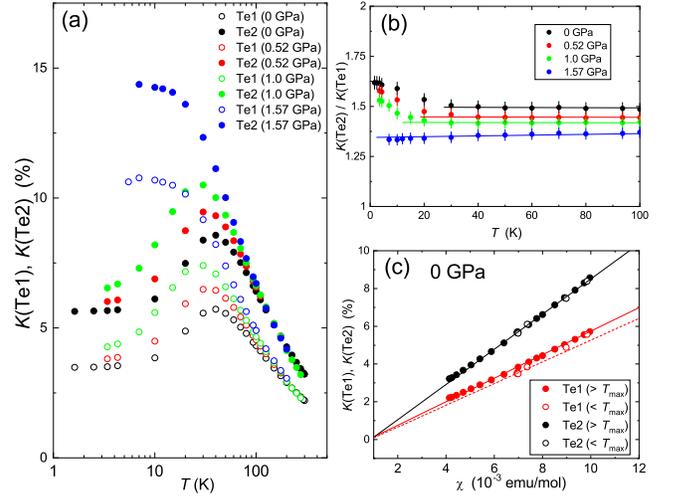} 
\caption{(a) $T$ dependences of $K$s for both Te1 and Te2 under different pressures. Open and closed symbols are for Te1 and Te2, respectively.
(b)  $T$ dependence of the ratio $K$(Te2)/$K$(Te1) under various pressures. The solid lines are guides for the eye. 
(c) $K$ versus magnetic susceptibility $\chi$ plots with $T$ as an implicit parameter. 
Solid and open symbols show the data above and below $T_{\rm max}$ = 35--40 K, respectively.    
The black solid line is a linear fit for Te2 for the whole temperature range.
The red solid line is a linear fit for Te1 for $T$ $\geq$ $T_{\rm max}$  and the red broken line represents a fit using the lowest temperature data while keeping a constant $y$-intercept corresponding to the temperature independent part of Knight shift. 
}
\label{fig:Fig2}
\end{figure}

\begin{figure*}[tb]
\includegraphics[width=180 mm]{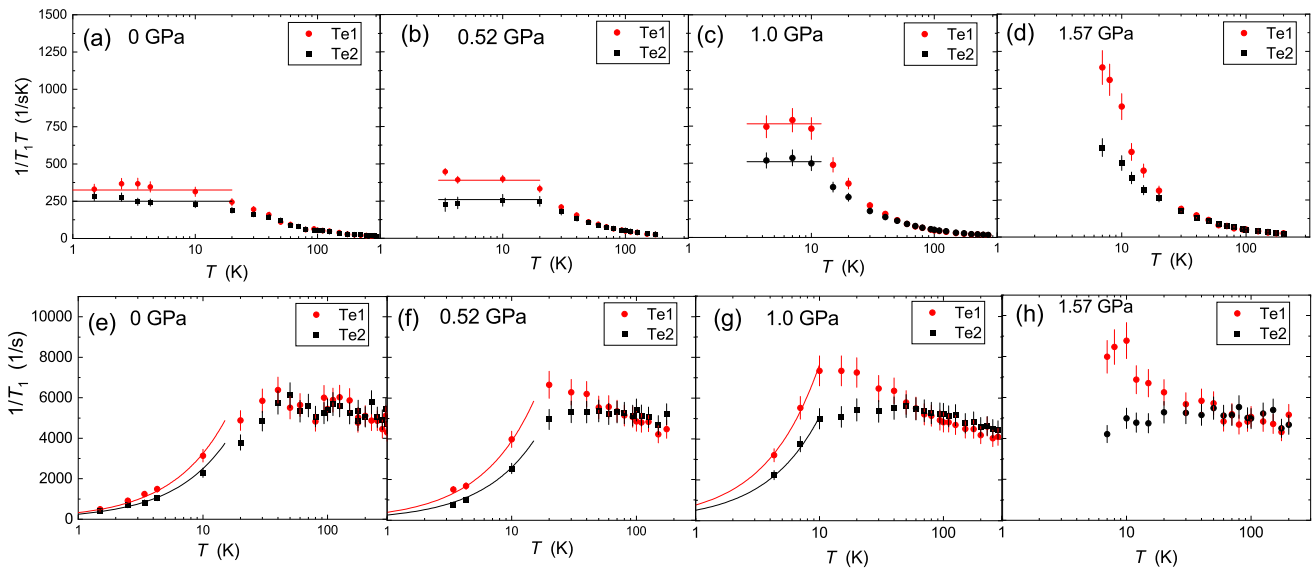} 
\caption{$T$ dependence of 1/$T_1T$ for Te1 (red circles) and Te2 (black circles) at $p$ = 0 (a), 0.5 (b), 1.0 (c) and 1.5 GPa (d). 
(e)-(h) $T$ dependence of 1/$T_1$ for Te1 (red circles) and Te2 (black circles) at $p$ = 0, 0.5, 1.0 and 1.5 GPa. 
The solid lines show 1/$T_1T$ = constant, expected for heavy-fermion state.  
}
\label{fig:Fig3}
\end{figure*}

     Figure\ \ref{fig:Fig1}(b) shows the $H$-swept $^{125}$Te-NMR spectra at 30 K under various pressures  ($p$ = 0--1.57 GPa) with $H$ parallel to the $b$ axis, where two lines corresponding to the two Te sites are observed.  
     The units of the horizontal axis are Knight shift $K$ defined by $K$ = (2$\pi$$f$/$\gamma_{\rm N}$ -- $H$)/$H$.       
     The two lines with lower and higher $K$ values have been assigned to Te1 and Te2, respectively \cite{Tokunaga2019}. 
     
     The temperature and pressure dependences of full width at half maximum (FWHM) of lines for Te1 and Te2 are shown in Fig. \ref{fig:Fig1}(c).
     At high $T$ above 200 K, the FWHMs are relatively small  $\sim$ 0.8 mT for both Te1 and Te2 at ambient pressure, showing a good sample quality. 
   With decreasing $T$, FWHM increases and shows a broad maximum around 35 K, similar to the $T$ dependence of magnetic susceptibility $\chi$ and also $K$ as shown in Fig.~\ref{fig:Fig2}(a). 
    This indicates that the $T$ dependence of FWHM reflects the $T$ dependence of $\chi$.  
    Similar $T$ dependence of FWHM can be observed with the lowered peak temperatures down to $\sim$20 K for $p$ = 0.52 and 1.0~GPa and $\sim$10 K at 1.57 GPa.
    The large enhancements of FWHM at higher $p$ at low $T$ suggest the increase of $\chi$, as actually observed in the $K$ data.


   The $T$ dependences of $^{125}$Te Knight shift for Te1 [$K$(Te1)] and Te2 [$K$(Te2)] sites determined by the peak positions of the NMR lines  are shown in Fig. \ref{fig:Fig2}(a)  for the measured pressures under $H$ $\parallel$ $b$.
    All $K$ values for Te2 are greater than those for Te1 due to the different hyperfine coupling constants ($A_{\rm hf}$).
    The $A_{\rm hf}$ values have been reported to be 34.1  and 51.8 kOe/$\mu_{\rm B}$ for Te1 and Te2, respectively, at ambient pressure \cite{Tokunaga2019}.
    
    At ambient pressure, $K$s for both Te sites show a similar temperature dependence with a broad maximum around $T_{\rm max}$ $\sim$ 35--40 K, similar to the  magnetic susceptibility data \cite{Ran2019, Aoki2019,Li2021}.
     With increasing $p$,  $T_{\rm max}$  shifts to lower temperatures to  $\sim$30 K at $p$ = 0.52 GPa and to $\sim$25 K at $p$ = 1.0 GPa. 
     These behaviors are consistent with the previously reported $p$-dependence of magnetic susceptibility \cite{Li2021} and the recent NMR data \cite{Kinjo2022}. 
      At $p$~=~1.57 GPa, both $K$s keep increasing with decreasing $T$ and level off below $\sim$15 K without showing a clear maximum. 
   We were able to measure the spectrum down to 5.5~K although the signal intensity becomes weak below 10 K due to the shortening of nuclear spin-spin relaxation time $T_2$. 
    However, we could not observe any signals at 4.2~K.     
    This could be due to the pressure-induced short-range magnetically ordered state whose onset temperature has been reported to be $\sim$5 K at $p$ = 1.57 GPa \cite{Ran_PRB2020,Thomas2020, Li2021}. 
     
    We notice that the $T$ dependences of $K$(Te1) and $K$(Te2) are slightly different below $\sim$30 K, a little bit lower than   $T_{\rm max}$ under $p$ $<$ 1.0 GPa.
   This can be seen in  Fig. \ref{fig:Fig2}(b) where the ratios of Knight shifts for the  two Te sites,  $K$(Te2)/$K$(Te1),  are plotted as  a function of $T$. 
    The ratio is nearly 1.48 at higher $T$ above $T_{\rm max}$  at ambient pressure and increases to 1.62 at  1.6 K. 
    This indicates that the $T$ dependences of $K$ for Te1 and Te2 are scaled above $\sim$30 K, but do not scale below 30 K.  
    Since the $T$ dependent part of $K$ is proportional to hyperfine coupling constant as $K(T)$ = $A_{\rm hf}$$\chi(T)$/$N_{\rm A}$$\mu_{\rm B}$  
where $N_{\rm A}$ is Avogadro’s number and $\mu_{\rm B}$  is Bohr magneton, this indicates that the ratio of the $A_{\rm hf}$ for Te1 and Te2 changes slightly  below $\sim$30 K. 
    To understand how the $A_{\rm hf}$ changes at low temperatures, we plotted $K$ as a function of the magnetic susceptibility. 
   As can be seen in Fig. \ref{fig:Fig2}(c), all the data points for Te2 are on the same straight line above and below 30 K, indicating no change in $A_{\rm hf}$ for this site.
   From the slope of the line,  $A_{\rm hf}$ is estimated to be 52.0 kOe/$\mu_{\rm B}$ which is very close to 51.8 kOe/$\mu_{\rm B}$ reported previously \cite{Tokunaga2019}.     
     On the other hand, although a clear linear relationship between $K$(Te1) and $\chi(T)$ can be seen above 30 K (shown by the solid red circles), the data points start deviating below the temperature as shown by the open red circles. 
    From the change in the slopes shown by solid and broken red lines in Fig. \ref{fig:Fig2}(c), we found the $A_{\rm hf}$ changes from 34.8 kOe/$\mu_{\rm B}$ above 30 K to 32.2 kOe/$\mu_{\rm B}$ at 1.6 K. 
    Thus we attribute the small increase in $K$(Te2)/$K$(Te1) below 30 K to the small reduction of $A_{\rm hf}$ for the Te1 site.
    Since 30 K is close to the crossover temperature below which Fermi-liquid (FL) behavior 1/$T_1T$ = constant is observed as shown in Fig. \ref{fig:Fig3}, the results suggest that $A_{\rm hf}$ changes slightly in the FL state in UTe$_2$.
    The similar $T$ dependence of the ratios can be seen at 0.52 and 1.0 GPa, but the deviation from the constant values at high temperatures starts at slightly lower temperatures of $\sim$~20~K for 0.52 GPa and $\sim$15 K for 1.0 GPa, respectively. 
    This is consistent with $T_1$ data where the crossover temperature to a low-temperature FL state decrease with increasing $p$. 
     It is interesting to point out that the ratios decrease with increasing $p$, indicating that the ratio of $A_{\rm hf}$ changes.  
    To determine the origin of the changes in the ratios under pressure, one needs $\chi$ data at the same pressure which are not available at present. 
    Further studies for the $\chi$ measurements are required to elucidate the origin.  
    At 1.57 GPa, the values of the ratio decrease slightly with $p$, but a clear upturn cannot be observed down to 5.5 K. 
    This may suggest no crossover to the FL state at 1.57~GPa down to $\sim$5.5 K.  
    It is also worth mentioning  that, since UTe$_2$ does not exhibit SC but has a magnetically ordered ground state under $p$ $>$ 1.5 GPa \cite{Thomas2020}, the unconventional SC state only appears in the FL state,  as has been pointed out in Ref. \cite{Kinjo2022}.

\begin{figure}[tb]
\includegraphics[width=\columnwidth]{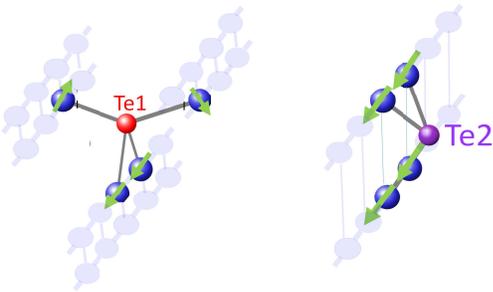} 
\caption{Images of the local spin fluctuations at the Te1 and Te2 sites.  
The dark blue circles represent the nearest neighbor U atoms for each Te site and the ladder structures along the $a$ axis are illustrated by the light blue circles. 
}
\label{fig:Fig4}
\end{figure}  

 Now we discuss magnetic fluctuations in UTe$_2$ based on the results of $T_1$ measurements for both Te sites.
   Figures \ref{fig:Fig3}(a)-\ref{fig:Fig3}(d) show the $T$ dependence of 1/$T_1T$ at various pressures.   
   At ambient pressure, 1/$T_1T$ for both the Te sites exhibits Curie-Weiss (CW)  behavior above $\sim$40 K.
    Below $\sim$40 K, due to the hybridization between the localized 5$f$-electron and conduction electron bands, 1/$T_1T$ exhibits nearly constant behavior, a characteristics of heavy-fermion states. 
    The solid lines represents the FL behaviors of 1/$T_1T$ = 325 and 250 1/sK for Te1 and Te2, respectively. 
    These results are consistent with the data reported previously \cite{Tokunaga2019}.
     The FL-crossover temperature defined as $T_{\rm FL}$ is also usually considered as Kondo temperature.

   Under pressure, the crossover temperature decreases down to  $T_{\rm FL}$ $\sim$ 20  and $\sim$ 10 K for  $p$ = 0.52 and 1.0 GPa, respectively. 
At the same time, the values of 1/$T_1T$~=~constant increase to 260 (390) 1/sK for Te2(Te1) at 0.52 GPa and to 510 (765) 1/sK for Te2(Te1) at 1.0 GPa, suggesting the increase of the density of states at the Fermi energy and/or the enhancement of electron correlations in the FL state. 
   The CW behavior of 1/$T_1T$ above $\sim$40 K is nearly independent of $p$, however, we observed a different $T$ dependence of 1/$T_1T$  for the two Te sites at low temperatures.
    With decreasing $T$ below $\sim$40 K,  1/$T_1T$ for Te1 is enhanced more than 1/$T_1T$ for Te2, which is clearly recognized  under $p$ = 1.57 GPa. 
    Even for the case of $p$ = 0.52 and 1.0 GPa, one can see difference in the $T$ dependence of 1/$T_1T$ below $\sim$40 K. 
    The different $T$ dependence between Te1 and Te2 is more clearly seen in the 1/$T_1$ vs. $T$ plots in Figs. \ref{fig:Fig3}(e)-\ref{fig:Fig3}(h).
    1/$T_1$ for Te2 is nearly $T$ independent at high temperatures and shows a FL behavior below $T_{\rm FL}$ $\sim$ 20~K and $\sim$ 10 K for  $p$~=~0.52 and 1.0 GPa, respectively. 
    1/$T_1$ for Te1 shows an enhancement where a clear difference in the $T$ dependence can be detected below $\sim$40 K down to $T_{\rm FL}$.   
    A similar difference in the $T$ dependence of 1/$T_1$ between Te1 and Te2 is also observed at $p$ = 1.57~GPa where the ground state is magnetic. 

    Those results indicate that Te1 and Te2 pick up different magnetic fluctuations.
    The  CW behavior in 1/$T_1T$ has been pointed out to be scaled with the $T$ dependence of the magnetic susceptibility under $H$ parallel to the $a$ axis (the magnetic easy axis), expected for FM spin fluctuations with $q$ = 0  \cite{Tokunaga2019}.  
The FM spin fluctuations were also suggested from nuclear spin-spin relaxation time $T_2$ measurements \cite {Tokunaga2019, Tokunaga2022}.    
    Therefore, the further enhancements of 1/$T_1T$ at the Te1 site in comparison with those at the Te2 site indicate an additional contribution of magnetic fluctuations with $q$ $\neq$ 0, AFM spin fluctuations, at  the Te1 site \cite{AFM_Te1}.
    These results suggest the existence of AFM and FM spin fluctuations in the paramagnetic state of UTe$_2$.
   It is noted that the AFM  spin fluctuations develop below around 40 K under pressure,  although the fluctuations were not clearly observed at ambient pressure.

    As mentioned above, Te2 is surrounded by  four U atoms forming a ladder and picks up magnetic fluctuations originating from each ladder.
    Thus our results indicate that FM spin fluctuations are dominant within the ladders. 
    On the other hand, since Te1 is surrounded by four U atoms belonging to three different ladders, it is possible to pick up magnetic fluctuations originating from not only intra-ladder but also inter-ladders magnetic couplings  as schematically shown in Fig. \ref{fig:Fig4}. 
   Therefore, the AFM spin fluctuations detected at Te1 are considered to be due to AFM coupling between the ladders in UTe$_2$.   
   These results are consistent with the NS measurements  where the intra-ladder magnetic interactions are ferromagnetic for the leg and the rung directions and the magnetic interactions between the ladders is antiferromagnetic \cite{Duan2020,Knafo2021}. 

   It is noted that the NS measurements detect the AFM spin fluctuations at ambient pressure while our NMR data do not show clear AFM spin fluctuation at $p$~=~0.    
   Although the reason for the difference is not clear at present, it may be possible to explain by taking into consideration the different energy scale between the two experimental techniques because NMR may not detect the magnetic fluctuations if those energy (frequency) are much higher than NMR frequency. 
    If this were the case, our NMR results might suggest that the fluctuation frequency of the AFM spin fluctuations gets lower with increasing pressure.  
Further NS experiments under pressure are required to clarify this.

   Finally it is interesting to point out that the energy scale of $\sim$40 K (below which the AFM spin fluctuations develop under $p$) has been detected in other experiments even at ambient $p$ such as the observation of peak in the magnetic excitation spectrum by inelastic NS measurements \cite {Butch2022}, the peak $T$ in the magnetic susceptibility, hybridization gap observed in scanning tunneling spectroscopy \cite{Jiao2020} and so on. 
   Therefore, the energy of $\sim$40~K could be considered as one of the characteristic energy scales in determining the physical properties of the heavy-fermion UTe$_2$. 


   In summary, we performed $^{125}$Te NMR measurements on UTe$_2$ under pressure up to 1.57 GPa.
    From the different temperature dependence of 1/$T_1T$ between the two different Te sites, Te1 and Te2,  we suggest the coexistence of AFM and FM spin fluctuations in  UTe$_2$.
   The FM spin fluctuations are considered to exist inside each ladder while the AFM spin fluctuations are assigned to originate from the inter-ladder magnetic interactions. 
    We point out that the FM spin fluctuations inside the ladders may play an important role in the appearance of the spin-triplet superconducting state in UTe$_2$, although AFM spin fluctuations start to develop below $\sim$40 K which are clearly observed under pressure.
    Further detailed investigations of the relationship between the magnitude of AFM/FM fluctuations and $T_{\rm c}$, the magnetic ordered states, and also about the superconducting properties under pressures are highly called for.

    The authors thank A. Sapkota, B. Li, M. Hirata, and R. Yamamoto for helpful discussions.  The research was supported by the U.S. Department of Energy (DOE), Office of Basic Energy Sciences, Division of Materials Sciences and Engineering. Ames Laboratory is operated for the U.S. DOE by Iowa State University under Contract No.~DE-AC02-07CH11358. The National High Magnetic Field Laboratory is supported by the National Science Foundation through NSF/DMR-1644779 and the State of Florida.


\begin{thebibliography}{10}
\bibitem{T1}     $1/T_1$ at each $T$ was determined by fitting the nuclear magnetization $M$ versus time $t$  using the exponential function $1-M(t)/M(\infty) = e^{-(t/T_{1})}$ for $^{125}$Te NMR,  where $M(t)$ and $M(\infty)$ are the nuclear magnetization at $t$ after the saturation and the equilibrium nuclear magnetization at $t$ $\rightarrow$ $\infty$, respectively.
\bibitem{AFM_Te1} It is noted that AFM spin fluctuations will not be canceled out at Te1 due to the local configuration with respect to the four U atoms forming the distorted tetrahedron as shown in Fig. 1(a), which is not a high symmetry point in the crystal. 
  In addition, if the AFM fluctuations are characterized by the incommensurate wave vector determined by the neutron scattering measurements, the AFM spin fluctuations will not be canceled out at the Te1 site.
\end{thebibliography}
\end{document}